\documentclass{elsart}
\usepackage{epsfig}
\usepackage{amssymb}
\begin{document}
\begin{frontmatter}
\title {$\eta$-meson photoproduction off protons and deuterons}
\author{C.~Deutsch-Sauermann\thanksref{email}},
\author{B.~Friman} and
\author{W.~N\"orenberg}  
\address{GSI, Planckstr.~1, D-64291 Darmstadt, Germany\\ and\\
Institut f\"ur Kernphysik, Technische Hochschule Darmstadt,\\
D-64289 Darmstadt, Germany}
\thanks[email]{e-mail: c.sauermann@gsi.de}

\begin{abstract}
We present a unitary and gauge-invariant model with coupled channels,
which provides a consistent description of pion photoproduction off
nucleons in the E$_{0+}$ channel and $\eta$-meson photoproduction off
protons and deuterons. An effective field theory with hadrons and
photons is constructed, which includes non-resonant Born terms as well
as the S$_{11}$(1535) and S$_{11}$(1650) baryon resonances. Due to the
coupling between the channels, the production of $\eta$-mesons is
strongly affected by the S$_{11}$(1650) although its direct coupling to
the $\eta$N channel is negligible. The rho- and omega-meson exchange
terms are important for achieving a consistent description of
both pion- and photon-induced reactions.
\end{abstract}
\end{frontmatter}

\section{Introduction and summary}
In recent years the physics of the $\eta$-meson has been studied
intensively. New measurements of $\eta$-production in hadronic and
heavy-ion collisions have been performed at SATURNE \cite{saturne}, Los
Alamos \cite{alamos}, Brookhaven \cite{clajus} and GSI \cite{gsi}.  The
availability of high-duty-factor electron accelerators, MAMI in Mainz
and ELSA in Bonn, has opened the possibility to perform precise
experiments with electromagnetic probes, e.~g.~ photon- and
electron-induced $\eta$-meson production off nucleons
\cite{krusche1,wilhelm,price} and nuclei \cite{kruschedeut,krusche2}.
The accurate measurements of electromagnetic processes provide strong
constraints on models for the elementary $\eta$-meson--hadron and
photon--hadron interactions, which are the basis for a theoretical
description of $\eta$-meson propagation in hadronic matter and
production in heavy-ion collisions. The theoretical interpretation of
the data has lead to the development of new models for photon-induced
$\eta$-meson production \cite{benn,mukhopadhyay,li}.

In this letter, we present a model for $\eta$-meson photoproduction
off nucleons and deuterons through the dominant E$_{0+}$
channel.  An effective field theory with hadrons and
photons is constructed, using a coupled-channels model \cite{paper}
for $\pi$N scattering and pion-induced $\eta$-production in the
S$_{11}$ channel as a starting point and coupling the electromagnetic
field to the hadrons in a gauge-invariant way.  As compared 
to existing models the following substantially new features 
are: 

\begin{itemize}
\item
{Resonance and background contributions are
taken into account in a strictly unitary way.  The 
consistent inclusion of final-state interactions between
the hadrons is important.}
\item
{In addition to the dominant S$_{11}$(1535) resonance, which couples strongly
to the $\eta$N channel, we also take the second
resonance S$_{11}$(1650) in the $\pi$N S$_{11}$ channel into account. We
find that the upper resonance is essential, although its small direct
coupling to the $\eta$N channel can be neglected.  This is due to the
coupling of the S$_{11}$(1650) to the $\eta$N channel through
intermediate $\pi$N states.  Because of the modifications due to the second
resonance, agreement with experimental data can be obtained only when
the $\rho$- and $\omega$-meson exchange terms are included.  These terms
are not needed in \cite{benn} because there
the S$_{11}$(1650) resonance is neglected.
The hadronic and electromagnetic coupling constants of the
S$_{11}$(1535) are, due to large interference effects, different from
those obtained in one-resonance models.}
\item
{The hadronic part of the model, which describes
the final-state interactions between the hadrons,
is determined exclusively by fitting
elastic $\pi$N scattering data. The consistency is checked by comparing
the cross section for the inelastic channel $\pi$N$\rightarrow \eta$N
with experiment.  Since the hadronic parameters are fixed, the
electromagnetic coupling constants are
uniquely determined by the photoproduction data. Thus we obtain
values for the helicity amplitudes of the resonances.
These are of great interest as a test for the quark model. 
Within our model we achieve a
consistent description of the pion photoproduction off nucleons in the
E$_{0+}$ channel and of the $\eta$-meson photoproduction off nucleons
and deuterons.}
\end{itemize}

\section{The model}
Let us briefly summarize the hadronic part of our model (for details
see \cite{paper}).  As already mentioned
there are two N$^*$ resonances in pion-nucleon scattering in the
S$_{11}$ channel at center-of-mass energies below $\sqrt{s}=$1.8~GeV
\cite{Hoeh}.  The lower one at 1535 MeV
(S$_{11}$(1535)) decays into $\pi$N, $\eta$N and $\pi\pi$N with the
branching ratios 35--55$\,\%$, 30--50$\,\%${} and 5--20$\,\%$,
respectively. The upper one at 1650 MeV (S$_{11}$(1650)) couples
strongly to the $\pi$N and $\pi\pi$N channels with branching ratios of
60--80$\,\%${} and 5--20$\,\%$, respectively, but only weakly to the
$\eta$N channel with a decay probability of approximately 1$\,\%$.

In order to describe these experimental facts, we include the
following interaction terms in the lagrangian,

\begin{eqnarray}\label{eq:a}
\mathcal{L}_I &=&-i  g_{\pi \mbox{\scriptsize NN}}
\overline{\Psi}_{\mbox{\scriptsize N}} \gamma_5 \vec{\tau} 
\Psi_{\mbox{\scriptsize
N}}\vec{\pi}
-g_{\sigma \mbox{\scriptsize NN}} \overline{\Psi}_{\mbox{\scriptsize N}}
\Psi_{\mbox{\scriptsize
N}}\sigma
-g_{\sigma\pi\pi} \vec{\pi}^2\sigma  \nonumber \\
& &- g_{\pi \mbox{\scriptsize N}\mbox{\scriptsize N}^*_1}\overline{\Psi}_
{\mbox{\scriptsize N}^*_1}\vec{\tau}\Psi_{\mbox{\scriptsize
N}}\vec{\pi}+\mbox{h.c.}
- g_{\pi \mbox{\scriptsize N}\mbox{\scriptsize N}^*_2}\overline{\Psi}_
{\mbox{\scriptsize N}^*_2}\vec{\tau}\Psi_{\mbox{\scriptsize
N}}\vec{\pi}+\mbox{h.c.} \\
& &- g_{\eta \mbox{\scriptsize N}\mbox{\scriptsize N}^*_1}\overline{\Psi}_
{\mbox{\scriptsize N}^*_1}\Psi_{\mbox{\scriptsize
N}}\eta+\mbox{h.c.} \nonumber \\
& &-i g_{\zeta \mbox{\scriptsize N}\mbox{\scriptsize N}^*_1}\overline{\Psi}
{\mbox{\scriptsize N}^*_1}\gamma_5\Psi_{\mbox{\scriptsize
N}}\zeta+\mbox{h.c.}
-i g_{\zeta \mbox{\scriptsize N}\mbox{\scriptsize N}^*_2}\overline{\Psi}_
{\mbox{\scriptsize N}^*_2}\gamma_5\Psi_{\mbox{\scriptsize
N}}\zeta +\mbox{h.c.}.\nonumber
\end{eqnarray}
The first line contains the interaction terms of the linear
sigma-model which incorporates chiral symmetry and describes
low-energy $\pi$N scattering.  In the second
line the $\pi$NN$^*${} interaction terms for the two N$^*$
resonances (N$^*_1\equiv$S$_{11}$(1535), N$^*_2\equiv$S$_{11}$(1650))
are given, whereas that in the third line is the $\eta$NN$^*_1${}
coupling.  We neglect the weak $\eta$NN and $\eta$NN$^*_2${}
couplings.  The effect of the physical two-pion continuum 
is parametrized by means
of an effective scalar field $\zeta${} of positive parity, mass
m$_{\zeta}$=400 MeV and zero width. The field $\zeta$ interacts 
through the $\zeta$NN$^*$ couplings, given in the last line
of~(\ref{eq:a}). The question why the $\sigma$- and the $\zeta$-fields
are not identical is discussed in \cite{paper}.

The Bethe-Salpeter equation, which couples the three open channels
$\pi$N, $\eta$N and $\zeta$N, is solved by using a $K$-matrix
approach, which guarantees a unitary $S$-matrix. 
We identify the $K$-matrix
elements with the diagrams shown in Fig.~1. 

\setlength{\unitlength}{1mm}
\begin{picture}(160,40)
\put(0,3)
{\epsfig{file=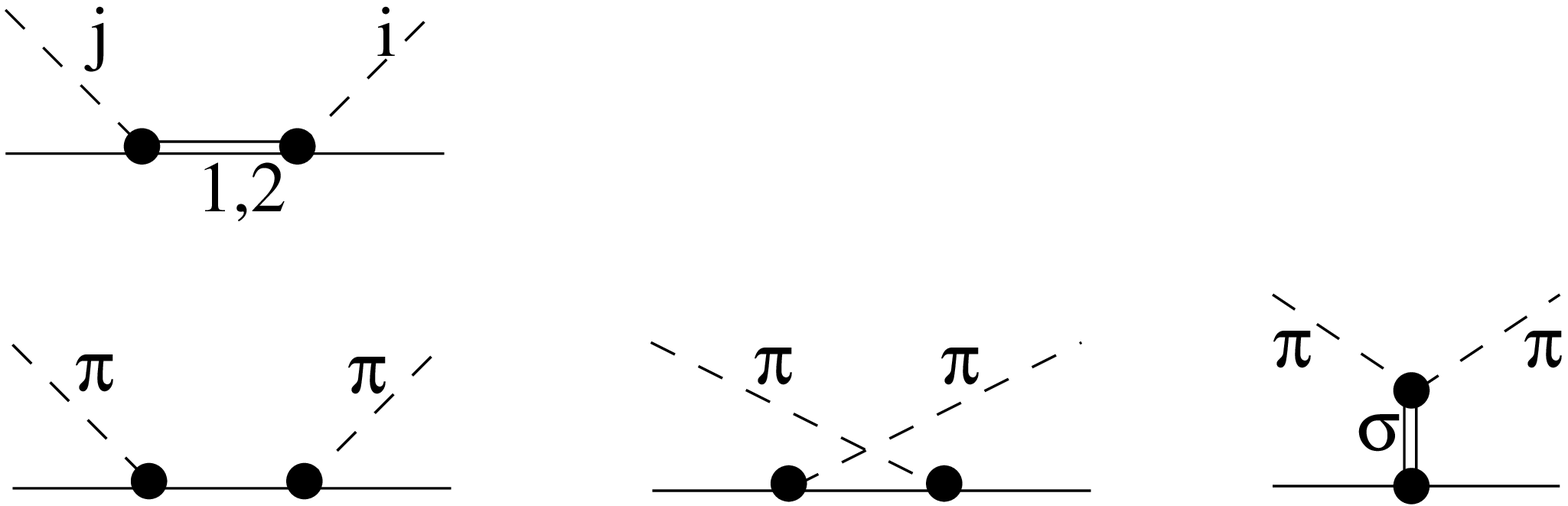,height=40mm}}
\end{picture}
\vspace*{-1cm}
\begin{center}
\parbox{13cm}{Fig.~1: Matrix elements $K_{j i}$: The first diagram shows the
contributions from the resonances, which contribute
in all three channels (i, j=$\pi,\eta,\zeta$; except $\eta$NN$_2^*$).
The last three diagrams correspond to the $\pi$N Born terms
which are included in the tree approximation to the linear sigma-model.}
\end{center}

In this approximation the $K$-matrix, and consequently also the $T$-matrix, is
free of divergences.  After introducing form factors at the
non-resonant Born terms, the coupling constants and resonance masses
are determined by fitting elastic $\pi$N scattering data, employing
the KA84 partial-wave analysis \cite{Hoeh}. The data are well
reproduced up to and including the energy region of the second
resonance \cite{paper}. 
A test of the consistency of the model is provided by the
cross section for the process $\pi^- + p \rightarrow \eta + n$.
The data for this process are well described by the model
without further adjustment of the parameters. It is important to
note the essential role played by the S$_{11}$(1650) resonance
not only in the elastic but also in the inelastic channel.
In previous calculations \cite{benn,bhal}
this resonance has not been included.

In this paper we extend the model by
introducing electromagnetic interactions. We stress that the hadronic
part of the model is not modified in any way.
To lowest order in the electromagnetic
interaction the $T$-matrix for the process
$\gamma$p$\rightarrow\eta$p is related to the $K$-matrix by
\begin{equation}\label{eq:b}
T_{\eta\gamma}=K_{\eta\gamma} - i \pi \sum_i T_{\eta i} \delta \left(
E-H_i\right)K_{i\gamma},
\end{equation}
where the sum is over all open channels, $i=\pi,\eta,\zeta$.  The
final-state interactions, described by $T_{\eta i}$, is entirely due to
strong interactions. In accordance with experiment \cite{krusche1}
we assume that the total cross section
for photoproduction of $\eta$-mesons is dominated by the S$_{11}$
channel and use our model for $\pi$N scattering and
pion-induced $\eta$-meson production in this channel to describe the
final-state interactions.

The resonance contributions to the $K$-matrix elements $K_{i\gamma}$
are identified with the first diagram shown in Fig.~2. The
$\gamma$NN$^*$ coupling is given by the lagrangian 
\begin{equation}\label{eq:c}
\mathcal{L}_{\gamma\mbox{\scriptsize N}\mbox{\scriptsize N}^*}=
\frac{-i e}{2\left(m_{\mbox{\scriptsize N}^*} + 
m_{\mbox{\scriptsize N}}\right)}\overline{\Psi}_{\mbox{\scriptsize N}^*}
\left(k_{\mbox{\scriptsize N}^*}^S+k_{\mbox{\scriptsize N}^*}^V 
\tau_3\right) \gamma^5 \sigma ^{\mu\nu}
\Psi_{\mbox{\scriptsize N}} F_{\mu\nu} + \mbox{h.c.},
\end{equation}
where $m_{\mbox{\scriptsize N}^*}$ and $m_{\mbox{\scriptsize N}}$ are
the resonance and nucleon masses. This interaction is
obviously gauge invariant, because it involves
only the electromagnetic field-strength tensor
$F_{\mu\nu}=\partial_{\nu} A_{\mu}-\partial_{\mu} A_{\nu}$.
Since the interaction has an isoscalar and an isovector
part, there are two new unknown coupling constants per resonance. 
They are proportional to the isoscalar and isovector helicity
amplitudes of the resonances.

\setlength{\unitlength}{1mm}
\begin{picture}(160,60)
\put(0,3)
{\epsfig{file=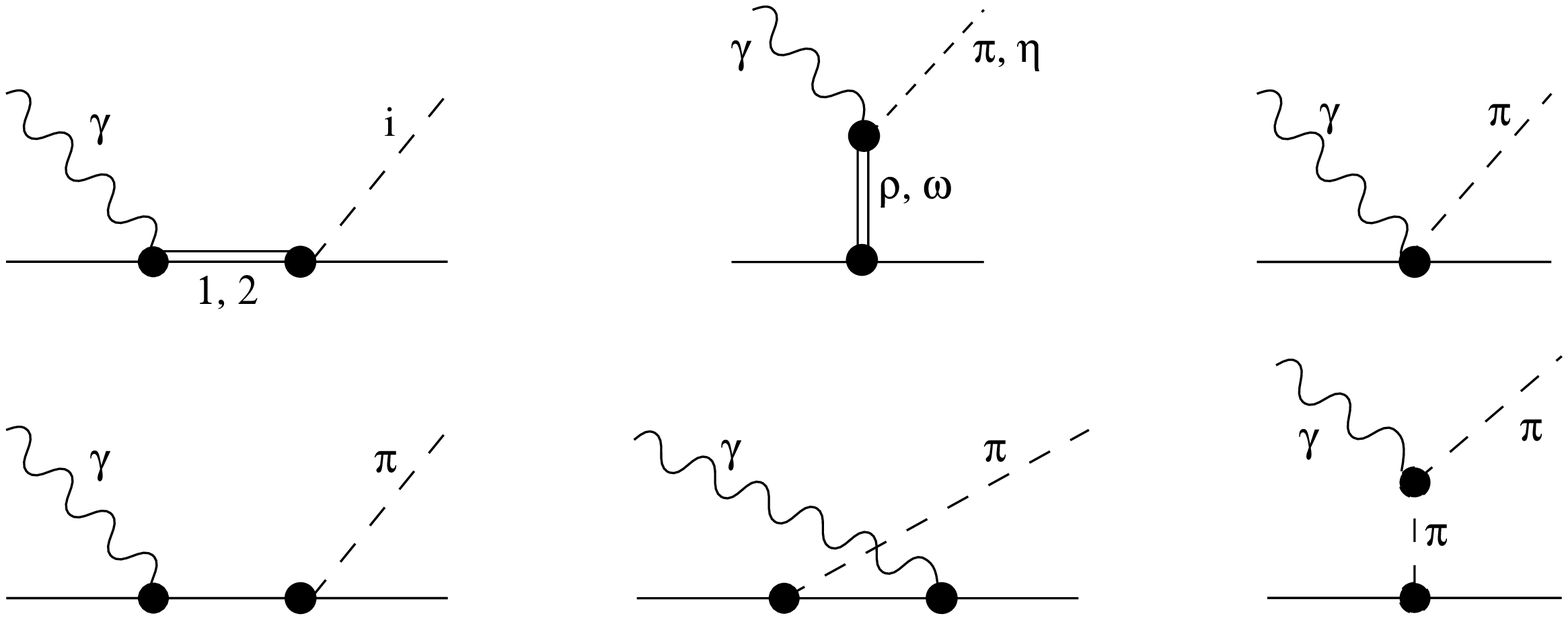,height=60mm}}
\end{picture}
\vspace*{-1cm}
\begin{center}
\parbox{13cm}{Fig.~2: Matrix elements $K_{i\gamma}$: 
The first diagram represents the
resonance processes which, 
except for the process $\gamma$N $\rightarrow$ N$_2^* \rightarrow \eta$N,
are non-zero (i=$\pi,\eta,\zeta$).  
The following diagram, showing $\rho$- and
$\omega$-meson exchange, contributes to $K_{\pi\gamma}$ as well as to
$K_{\eta\gamma}$, whereas the remaining ones correspond to $\pi$N Born
terms.}
\end{center}

Furthermore, the $\rho$- and $\omega$-meson exchange processes, represented 
by the second diagram in Fig.~2, give important
contributions to $K_{\pi\gamma}$ as well as to $K_{\eta\gamma}$.
The corresponding lagrangian reads
\begin{eqnarray}\label{eq:d}
\mathcal{L}_V
&=&-f_{\omega{\mbox{\scriptsize NN}}}\overline{\Psi}_{\mbox{\scriptsize N}}
\gamma^\mu \Psi_{\mbox{\scriptsize N}}  \omega_\mu
-f_{\rho {\mbox{\scriptsize NN}}}\overline{\Psi}_{\mbox{\scriptsize N}}
\gamma^\mu \vec{\tau} \Psi_{\mbox{\scriptsize N}} \vec{\rho}_\mu
+\frac{f_{\rho {\mbox{\scriptsize NN}}}}
{2 M}\kappa_\rho\overline{\Psi}_{\mbox{\scriptsize N}}
\sigma ^{\mu\nu} \vec{\tau}
\Psi_{\mbox{\scriptsize N}}\partial_{\nu}\vec{\rho}_{\mu} \nonumber \\
& &+\frac{e\lambda_{\omega\eta\gamma}}{4 m_\eta}
\epsilon_{\alpha\beta\gamma\delta}F^{\alpha\beta}\omega^{\gamma\delta}
\eta
+\frac{e\lambda_{\rho\eta\gamma}}{4 m_\eta}
\epsilon_{\alpha\beta\gamma\delta}F^{\alpha\beta}\rho_0^{\gamma\delta}
\eta \\
& &+\frac{e\lambda_{\omega\pi\gamma}}{4 m_\pi}
\epsilon_{\alpha\beta\gamma\delta}F^{\alpha\beta}\omega^{\gamma\delta}
\pi_0
+\frac{e\lambda_{\rho\pi\gamma}}{4 m_\pi}
\epsilon_{\alpha\beta\gamma\delta}F^{\alpha\beta}\vec{\rho}\,^{\gamma\delta}
\cdot \vec{\pi} \nonumber .
\end{eqnarray}
Here, $\omega^{\gamma\delta}$ and $\rho^{\gamma\delta}$ are the field
strength tensors of $\omega$- and $\rho$-meson, respectively. The first
line contains the interaction of $\omega$- and $\rho$-meson with the
nucleon, where the corresponding coupling constants are well known
\cite{Hoe2}: $f_{\rho \mbox{\scriptsize{NN}}}=2.52, f_{\omega
\mbox{\scriptsize{NN}}}=3f_{\rho\mbox{\scriptsize{NN}}}$ and
$\kappa_{\rho}=6.6$. 
We neglect the small coupling
of the $\omega$-meson to the nucleon
anomalous magnetic moment. The interactions which are responsible for the
decays of $\omega$ and $\rho$ mesons into a photon and an $\eta$- or
$\pi$-meson are given in the two bottom lines. The constants
$\lambda_{\omega\eta\gamma}=0.329$, $\lambda_{\rho\eta\gamma} = 1.02$
and $\lambda_{\omega\pi\gamma}=0.32$ are determined by the
experimentally well known decay widths. The uncertainty in the branching
ratio for the decay of a neutral $\rho$-meson into $\pi^0 \gamma$ is
rather large, while the corresponding decay for the $\omega$-meson is
quite well known. We therefore use the SU(2) relation
$\lambda_{\rho\pi\gamma} =\lambda_{\omega\pi\gamma}/3 $ to determine the
$\rho\pi\gamma$ coupling constant. The resulting value is compatible
with the measured branching ratio.

We also include non-resonant Born terms in $K_{\pi\gamma}$ in order to be
consistent with the interaction terms of the linear sigma-model and to
fulfil the low-energy theorems of pion photoproduction.  The
corresponding interaction terms are obtained by minimal coupling of
the photon to the lagrangian of the linear sigma-model.
This generates a coupling of the photon to the electromagnetic 
currents of the nucleon and the pion, 
and consequently in tree approximation the diagrams
in the second row of Fig.~2. We obtain the lagrangian
\begin{eqnarray}\label{eq:e}
\mathcal{L}_{\gamma \mbox{\scriptsize N}\mbox{\scriptsize N}} +
\mathcal{L}_{\gamma \pi\pi}&=& -e\overline{\Psi}_{\mbox{\scriptsize N}}
\frac{\left(1+\tau_3\right)}{2} \gamma^\mu A_\mu
\Psi_{\mbox{\scriptsize N}} + \frac{1}{2}
\partial_\mu\vec{\pi} i e A^\mu T_3 \vec{\pi}+ \frac{1}{2}i e A_\mu
T_3\vec{\pi} \partial^\mu\vec{\pi} \nonumber \\ & &+\frac{e}{4
m_{\mbox{\scriptsize N}}}\overline{\Psi}_{\mbox{\scriptsize N}}
\left(k^S+k^V \tau_3\right)\sigma ^{\mu\nu} \Psi_{\mbox{\scriptsize
N}} F_{\mu\nu},
\end{eqnarray}
where we have added the coupling to the anomalous magnetic moment.
Here $k^S$ and $k^V$ are the isoscalar and isovector part of the
anomalous magnetic moment of the nucleon, $k^S=-0.06$ and $k^V=1.85$.

Since the linear sigma-model is chirally invariant, PCAC 
and consequently the low-energy theorems of pion photoproduction are
satisfied. However, as shown in \cite{davidson}, 
the low-energy theorems are not satisfied on the tree level but only
once one-loop diagrams are included. 
In order to satisfy the low-energy theorems in the
absence of loop diagrams an additional contact interaction of the form
\begin{equation}\label{eq:f}
\mathcal{L}^c= \frac{i e g_{\pi\mbox{\scriptsize{NN}}}}{8
m_{\mbox{\scriptsize N}}^2}\overline{\Psi}_{\mbox{\scriptsize N}}
\gamma^5 \{\vec{\tau},k^S+k^V\tau_3\}\vec{\pi}\sigma ^{\mu\nu}
\Psi_{\mbox{\scriptsize N}} F_{\mu\nu}
\end{equation}
has to be added, which contains an anticommutator of nucleon isospin
matrices.  One can construct this form by starting from a
lagrangian with pseudovector $\pi$NN interaction, coupling the photon in a
minimal way, adding a coupling to the anomalous magnetic moment of the
nucleon and then performing a chiral rotation of the baryon
fields \cite{fubini,drechsel}.
Neglecting higher-order interaction terms, one ends up with a
model with pseudoscalar $\pi$NN interaction plus the additional
interaction term (\ref{eq:f}), which
satisfies the low-energy theorems.  We note that this contact term, shown in
Fig.~2, should be distinguished from the Kroll-Ruderman term,
which is obtained by coupling the photon in a minimal way to a
pseudovector $\pi$NN-interaction.

Now the essential contributions are accounted for. The contributions
of other resonance or meson exchange terms
are expected to be small 
either due to small decay widths into the relevant channels
or because of large masses in intermediate states.

We introduce form factors for the contact term and the vector-meson
exchange contributions.  Gauge invariance implies large cancellations
between the form factor at the $\pi$NN-vertex 
and additional terms generated by coupling
the electromagnetic field to the form factor in a minimal way.
The net effect is that for the current coupling,
the diagrams in the bottom row of Fig.~2 
should be computed without form factors because the form factor
corrections to these diagrams are cancelled by other contributions
\cite{ohta,gross,naus}.

\section{Results}
The model is now defined and the electromagnetic
parameters, namely the isoscalar and isovector couplings of the photon
to the two resonances as well as the cutoffs,
should be determined. The ideal way to proceed would
be to fit the parameters to the E$_{0+}$-amplitude of pion
photoproduction and then predict the total cross section for
photoinduced $\eta$-production. However, in view of the fact that there
are substantial deviations between different existing partial-wave
analyses for pion photoproduction this is not a useful way to proceed. 
Therefore we choose a more pragmatic approach and
search for a optimal description of both $\pi$-
and $\eta$-photoproduction. 
Considering the uncertainty in the analyses, we obtain 
reasonable agreement with the data on the
E$_{0+}$-amplitude of pion photoproduction in all isospin channels
\cite{arndt} for a parameter set which also describes
the total cross section of $\eta$-production off protons and
deuterons \cite{thesis}.  We find the following helicity amplitudes (in
units of $10^{-3}$ GeV$^{-\frac{1}{2}}$): $A_{1/2}^p$=102,
$A_{1/2}^n$=--82 for the S$_{11}$(1535) and $A_{1/2}^p$=83 and
$A_{1/2}^n$=--24 for the S$_{11}$(1650), which correspond to
$k_{\mbox{\scriptsize N}^*(1535)}^S=0.09, k_{\mbox{\scriptsize
N}^*(1535)}^V=0.84, k_{\mbox{\scriptsize N}^*(1650)}^S=0.25$ and
$k_{\mbox{\scriptsize N}^*(1650)}^V=0.45$.  The resulting total
cross section of the process $\gamma$+p$\rightarrow\eta$+p is shown by
the solid line in Fig.~3, together with data
from \cite{krusche1,wilhelm}.  

 \setlength{\unitlength}{1mm}
\begin{picture}(160,80)
\put(0,0)
{\epsfig{file=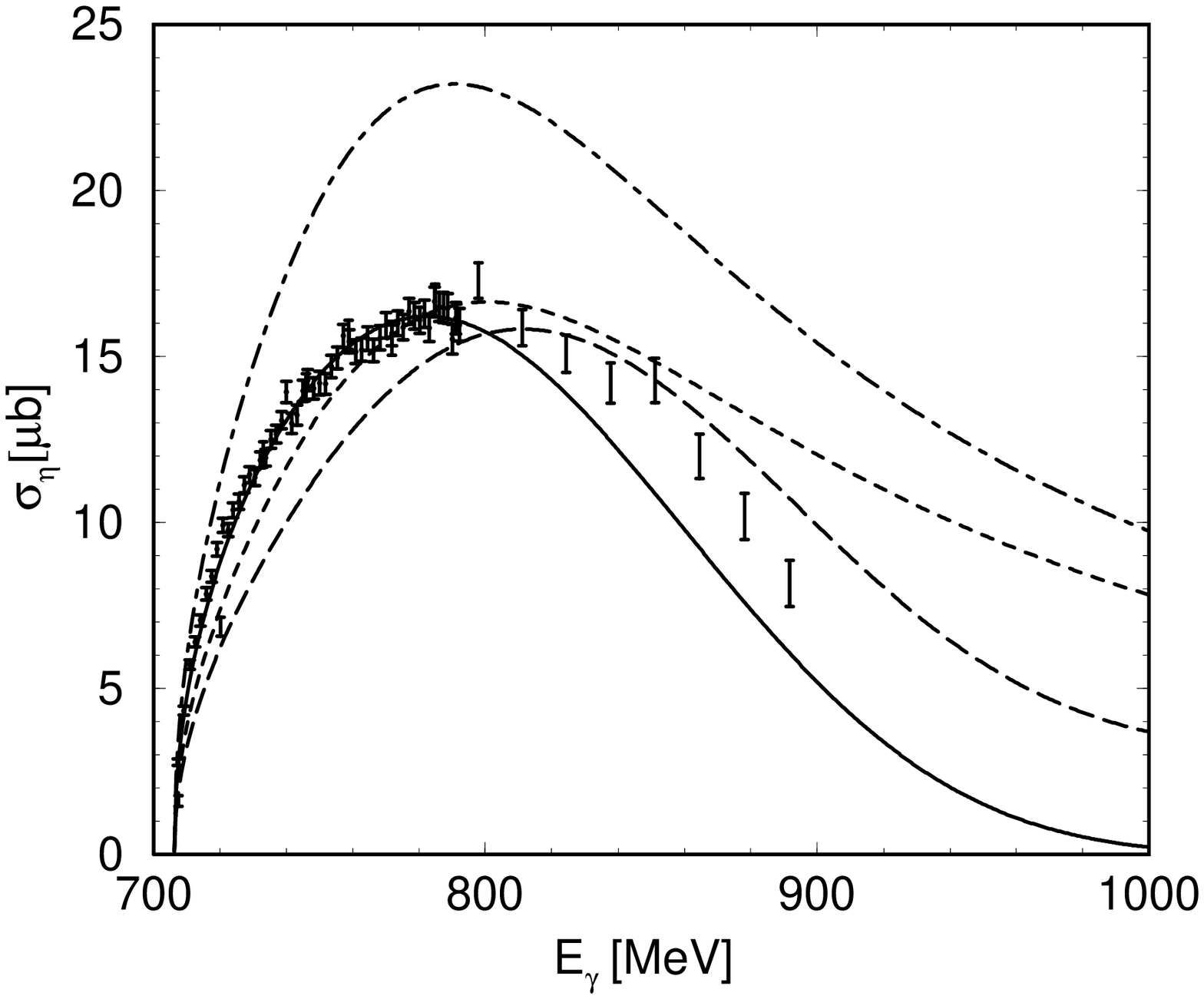,height=80mm}}
\end{picture}
\begin{center}
\parbox{13cm}{Fig.~3: 
Total cross sections for the photoproduction of $\eta$-mesons
off protons (solid line). The data are taken from
\cite{krusche1,wilhelm}. See text for further details.}
\end{center}

It is instructive to examine the contributions of the different
processes to the cross section.
The short-dashed line shows the cross section
when all processes except those involving the S$_{11}$(1535)
are neglected. Accidentally this is relatively close to 
the data.  The dot-dashed line is obtained by 
including also the non-resonant pion-nucleon Born terms, while the long-dashed
line is found by adding also the second resonance.
Thus, in the calculation corresponding to the long-dashed line only
the vector-meson exchange terms are neglected. Obviously the
presence of the second resonance leads to destructive interference and a
shift of the peak towards higher energies compared to a model which
includes only one resonance.  This shift is compensated by the vector
meson exchange contributions. Hence, when the second resonance is
included one cannot describe simultaneously both the 
$\pi$N $\rightarrow$ $\eta$N
and  $\gamma$N $\rightarrow$ $\eta$N data without vector-meson exchange
contributions. In other models, where
the second resonance is ignored, the vector-meson
exchange contributions are not needed \cite{benn}.  Although the direct
coupling of the S$_{11}$(1650) to the $\eta$N channel is very weak and
consequently neglected in our model, this resonance plays an important
role in the photoproduction of $\eta$-mesons, due
to the coupling to the $\eta$N-channel via intermediate
$\pi$N states. This effect can never be described
in tree level calculations \cite{mukhopadhyay}.
 
The total cross section for $\eta$-meson photoproduction 
off deuterons is computed in the impulse approximation,
i.~e. we assume that the elementary meson-production amplitudes from
the two nucleons in their bound state 
can be added to form the production amplitude from
the deuteron.  The corrections due to final-state interactions
between the nucleons as well as between the produced meson and the
spectator nucleon are expected to be 
small\footnote{except near threshold where the NN cross section is large
(see below)}, because
of the large distance between the two nucleons in the deuteron
\cite{vijay}. 
Since the nucleon which is hit by the photon is not on its mass shell in the
deuteron, we must use half-off-shell elementary amplitudes here.
In the impulse approximation the other nucleon is a spectator,
which must be on its mass shell. Thus,
the energy $E_{\mbox{\scriptsize N}}$ of the off-shell
nucleon is, for a given value of its momentum 
$\vec{p}_{\mbox{\scriptsize N}}$, determined by energy
and momentum conservation.
The initial $K$-matrix element $K_{i\gamma}$ is computed with an
off-shell nucleon with energy $E_{\mbox{\scriptsize N}}$ 
and momentum $\vec{p}_{\mbox{\scriptsize N}}$ and an on-shell photon
with energy $E_\gamma$ in the initial state and on-shell particles
in the final state.
The subsequent interactions are treated in the $K$-matrix
approximation, with on-shell propagation
at an invariant mass squared $s=(E_{\mbox{\scriptsize N}}+E_\gamma)^2
-(\vec{p}_{\mbox{\scriptsize N}}+\vec{p}_\gamma)^2$.

The $T$-matrix for photoproduction off the deuteron
is obtained by multiplying with the deuteron wave
function $\Psi_D(p_{\mbox{\scriptsize N}})$ of
the Paris potential \cite{paris}. The resulting
 cross section is in good agreement with data
(cf.~Fig.~4). 
Note that the momentum distribution of the nucleons in the
deuteron has an important effect on the cross section.
At small photon energies one 
observes the so-called subthreshold production. 
At the peak, the deuteron cross section is much smaller
than twice the proton cross section, whereas
at $E_{\gamma}\approx 850$ MeV their ratio is about two, in
agreement with experiment \cite{kruschedeut,bacci}.
This has been interpreted in terms of different
energy dependencies of the neutron and proton cross sections:
At low energies $\sigma_n/\sigma_p\approx 2/3$, while at energies
above $E_\gamma\approx$ 850 MeV a ratio exceeding
unity was found, implicitly assuming that the nucleons in the deuteron are
at rest. In our model, which is consistent with the data,
the ratio $\sigma_n/\sigma_p$
is almost constant for $E_\gamma \leq 900$ MeV and increases for larger 
energies, reaching unity at $E_\gamma\approx 950$ MeV.
Thus, the smearing in energy due to the momentum distribution
plays a crucial role here, and invalidates the naive interpretation
of the deuteron cross section.

\setlength{\unitlength}{1mm}
\begin{picture}(160,80)
\put(0,0)
{\epsfig{file=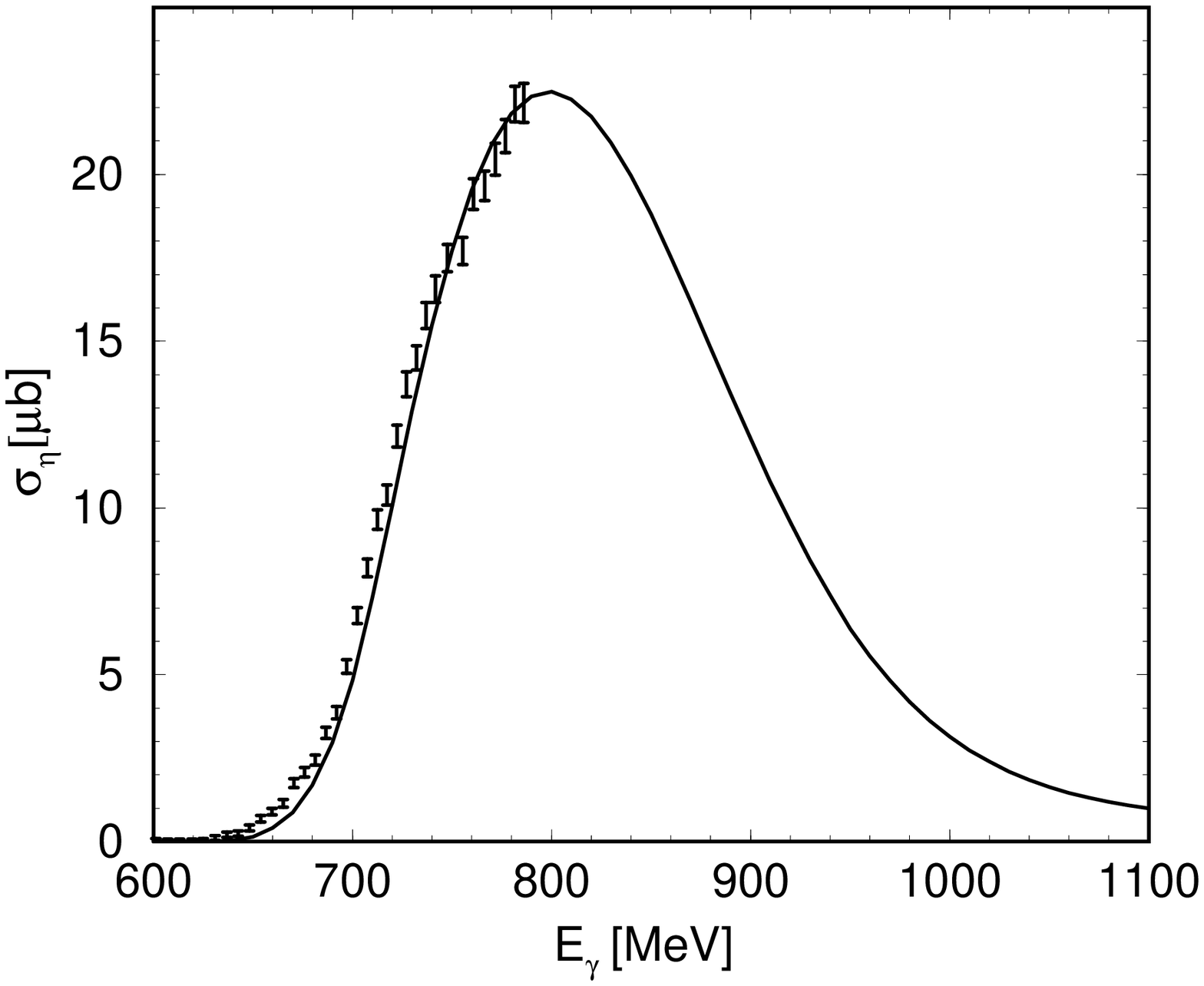,height=80mm}}
\end{picture}
\begin{center}
\parbox{13cm}{Fig.~4: 
Total cross section for the photoproduction of $\eta$-mesons 
off the deuteron. Data are taken from \cite{kruschedeut}.}
\end{center}

In order to explore the sensitivity of the cross section to
uncertainties in the deuteron wave function we repeated
the calculation with the wave function of
the Bonn potential \cite{bonn}. The results are indistinguishable, 
since the wave functions differ significantly only
at relative nucleon momenta higher than 400 MeV, which
play a negligible role here.

Due to the dominant isovector character of the S$_{11}$(1535) helicity
amplitudes one might expect strong destructive interference between
the two elementary $\gamma$N $\rightarrow \eta$N amplitudes, 
where in the one case the photon hits the proton
and in the other the neutron.
However, the interference term is negligible because of the
following reason.
The participating nucleon in one of the two interfering matrix
elements is a spectator in the other one. Since the
final states of both matrix elements have to be identical, this means
that, due to the large momentum transfer to the participating nucleon,
the interference term probes the deuteron wave function at high
relative momenta, where it is very small.

Finally, we note that near threshold the model underestimates 
the data by $\sim 50 \%$.
The discrepancy may be due to proton--neutron final-state interactions
\cite{vijay}. A calculation including this effect is in progress.

\section{Conclusion}
The $\eta$-meson photoproduction off nucleons is
described within a unitary and gauge-invariant model. Elastic $\pi$N
scattering data are used to determine the hadronic parameters of the
model.  The predicted cross section for pion-induced $\eta$-production
agrees well with the data.  We obtain a consistent description of
the photoproduction of $\eta$-mesons and pions 
which satisfies the low-energy theorems for the latter reaction. 
In all these processes 
not only the S$_{11}$(1535)-resonance but also the
S$_{11}$(1650)-resonance plays an important role.
We also find that the $\rho$- and $\omega$-exchange 
processes play a crucial role in the photoproduction of $\eta$-mesons.

Estimates of the $\eta$NN coupling constant in the framework of a
nonlinear \linebreak 
SU(3)$\times$SU(3) sigma-model with mesons and baryons
imply that the contribution
of the $\eta$-nucleon pole terms to the photoproduction of $\eta$-mesons
is small and can be neglected \cite{weise}. Thus, our model provides a
consistent description of $\eta$-meson production, with all
essential contributions included.

\section*{Acknowledgment}
We thank B.~Krusche, V.~Metag, V.~R.~Pandharipande, D.~O.~Riska,
M.~Soyeur, H.~Str{\"o}her and W.~Weise for valuable discussions.

\end{document}